\documentclass{aamas2016}

\begin{document}

% In the original styles from ACM, you would have needed to
% add meta-info here. This is not necessary for AAMAS 2015  as
% the complete copyright information is generated by the cls-files.

\title{Wheeled Robots playing Chain Catch: Strategies and Evaluation}

% AUTHORS

% For initial submission, do not give author names, but the
% tracking number, instead, as the review process is blind.


%\numberofauthors{8} %  in this sample file, there are a *total*

\numberofauthors{2}

\author{
% 1st. author
\alignauthor
%Paper number 50 \\
% Paper XXX
Garima Agrawal\\
       \affaddr{International Institute of Information Technology}\\
       \affaddr{Hyderabad, India}\\
%     \affaddr{Wallamaloo, New Zealand}\\
       \email{garima.agrawal@research.iiit.ac.in}
% 2nd. author
\alignauthor
Kamalakar Karlapalem\\
       \affaddr{IIIT Hyderabad/IIT Gandhinagar}\\
      \affaddr{ India}\\
%        \affaddr{Dublin, Ohio 43017-6221}\\
       \email{kamal@iiit.ac.in}
       }	
% 3rd. author
%\alignauthor Lars Th{\o}rv{\"a}ld\titlenote{This author is the one who did all the really hard work.}\\
%       \affaddr{The Th{\o}rv{\"a}ld Group}\\
%       \affaddr{1 Th{\o}rv{\"a}ld Circle}\\
%       \affaddr{Hekla, Iceland}\\
%       \email{larst@affiliation.org}

%\and  % use '\and' if you need 'another row' of author names

% 4th. author
%\alignauthor Lawrence P. Leipuner\\
%       \affaddr{Brookhaven Laboratories}\\
%       \affaddr{Brookhaven National Lab}\\
%       \affaddr{P.O. Box 5000}\\
%       \email{lleipuner@researchlabs.org}

% 5th. author
%\alignauthor Sean Fogarty\\
%       \affaddr{NASA Ames Research Center}\\
%       \affaddr{Moffett Field}\\
%       \affaddr{California 94035}\\
%       \email{fogartys@amesres.org}

% 6th. author
%\alignauthor Charles Palmer\\
%       \affaddr{Palmer Research Laboratories}\\
%      \affaddr{8600 Datapoint Drive}\\
%       \affaddr{San Antonio, Texas 78229}\\
%       \email{cpalmer@prl.com}

%\and

%% 7th. author
%\alignauthor Lawrence P. Leipuner\\
%       \affaddr{Brookhaven Laboratories}\\
%       \affaddr{Brookhaven National Lab}\\
%       \affaddr{P.O. Box 5000}\\
%       \email{lleipuner@researchlabs.org}

%% 8th. author
%\alignauthor Sean Fogarty\\
%       \affaddr{NASA Ames Research Center}\\
%       \affaddr{Moffett Field}\\
%       \affaddr{California 94035}\\
%       \email{fogartys@amesres.org}

%% 9th. author
%\alignauthor Charles Palmer\\
%       \affaddr{Palmer Research Laboratories}\\
%       \affaddr{8600 Datapoint Drive}\\
%       \affaddr{San Antonio, Texas 78229}\\
%       \email{cpalmer@prl.com}

%}

\maketitle

\begin{abstract}
Robots playing games that humans are adept in is a challenge. We studied robotic agents playing Chain Catch game as a Multi-Agent System (MAS). Our game starts with a traditional Catch game similar to Pursuit evasion, and further extends it to form a growing chain of predator agents to chase remaining preys. Hence Chain Catch is a combination of two challenges - pursuit domain and robotic chain formation. These are games that require team of robotic agents to cooperate among themselves and to compete with other group of agents through quick decision making. In this paper, we present a Chain Catch simulator that allows us to incorporate game rules, design strategies and simulate the game play. We developed  cost model driven strategies for each of Escapee, Catcher and Chain. Our results show that Sliding slope strategy is the best strategy for Escapees whereas Tagging method is the best method for chain$'$s movement in Chain Catch. We also use production quality robots to implement the game play in a physical environment and analyze game strategies on real robots. Our real robots implementation in different scenarios shows that game strategies work as expected and a complete chain formation takes place successfully in each game.
\end{abstract}

% Note that the category section should be completed after reference to the ACM Computing Classification Scheme available at
% http://www.acm.org/about/class/1998/.


%A category including the fourth, optional field follows...
%\category{D.2.8}{Software Engineering}{Metrics}[complexity measures, performance measures]

%General terms should be selected from the following 16 terms: Algorithms, Management, Measurement, Documentation, Performance, Design, Economics, Reliability, Experimentation, Security, Human Factors, Standardization, Languages, Theory, Legal Aspects, Verification.


%Keywords are your own choice of terms you would like the paper to be indexed by.


\section{Introduction}
We implement robotic agents playing Chain Catch, which is a common multi-player playground game that requires strategic decision making and cooperation among chain members to stay together (as a chain) while catching another player whereas other players to compete with chain to escape from getting caught. Simulating robot games like Robo-soccer and Robot pursuit evasion games have been a topic of extensive research in the field of Multi-Robot systems \cite{31,36,57}. Our game starts as simple Catch-Catch or ``tag'' game that falls under pursuit domain problems. In our Chain Catch game (i) the Catcher Catches one of the Escapees, (ii) the Catcher and caught Escapee form a chain to Catch other Escapees and (iii) step (ii) is repeated until all Escapees are caught and become part of one chain. Chain Catch requires complex and efficient strategies to counter the Catcher or chain. The game also requires us to develop techniques for robotic chain formation and movement suitable in game scenario. Our Chain Catch agents are autonomous and compute their strategy in a decentralized  manner. 

\subsection{The Chain Catch Game setup}
The world of the Robo Chain Catch game is a grid of size \textit{Width*Height} (can be varied).  Each cell in the grid is considered as one unit and has distinct position given by its \textit{x} and \textit{y} coordinates. The world is surrounded by four boundary edges that can be seen as walls, as the agents need to restrict their motion within these boundaries. Two agents cannot occupy the same cell. A Chain Catch agent is defined by three parameters, (\textit{x}, \textit{y}, \textit{CatchMode}). Following are the game rules, terms and parameters used throughout this paper-\\
\textbf{Catcher} Agent, that is assigned the role of chasing and catching other agents is a Catcher. \\
\textbf{Escapee} Agents, that try to evade from the Catcher or chain in the game are Escapees.\\
\textbf{Chain member} Each agent part of the chain is called a Chain member.\\
\textbf{Agent Diameter} Each Agent has a specific diameter which is specified in the units of cell. \\
\textbf{Visible range} All the agents can see all other agents in the field. \\
\textbf{Legal Moves} The agents are allowed to move to their eight adjacent cells, including diagonal cells.  \\
\textbf{Catch} A Catch is said to happen between a catcher or chain member and an escapee if distance between them is less than or equal to Agent Diameter.\\
\textbf{Chain constraints} Arrangement of robots is considered as a chain when each member of chain has two neighbours - (one to its left and one to its right) at a minimum distance $r_1$ (Agent Diametre) and maximum $r_2$ (2*Agent Diametre) apart. Two ends of chain should not meet each other and should have one neighbour each. If the chain breaks  (not binding to chain constraints) in between the game, a Catch occurred during that time is rejected.\\
\textbf{Game Over} The game gets over when all the agents in game become part of one chain. However, in practice, robots get discharged after a certain time. Therefore, we limit total game time in terms of maximum number of steps taken by an agent which is based on size of arena.

\subsection{Related Work}

 There is work done on simulating robot games as Multi-agent systems like Robo-soccer \cite{36,29,51} and pursuit domain games such as man and lion game \cite{47,41} and traditional pursuit evasion \cite{31,57,57}. However, there has not been any work in direction of implementing Chain Catch as a multi-robot system. Korf suggested a standard solution to the pursuit problem \cite{34} using the concept of attractive and repulsive forces. Our Escapees$’$ naive strategy is inspired from Korf$’$s solution to prey$’$s motion. However, there are more than one number of Escapees (preys) in our game, hence they require to have explicit cooperation among themselves to counter and disturb the Catcher/Chain strategy. 
 
 Lion and man problem \cite{47} is also another game from pursuit domain. Gale does not consider boundary condition and allows alternate turns of players, Whereas in our game all the agents move simultaneously restricted to four boundaries. Game theoretical approaches can also be used for prey-predator games \cite{33}.  But however this approach is centralized, as there has to be a central server who does the calculation of the payoff function for all the
possible strategies and then intimate predators with corresponding move that leads to capture the prey.

In the chain formation behavior, robots have to position themselves in order to connect to their two neighbours. In Mead et al, control of robot formation shapes is achieved by treating each robot as a cell in a cellular automaton, where local interactions between robots result in a global organization \cite{39}. 
 Maxim used virtual physics-based design to form chains of robots \cite{37}. In their application, first robot remains stationary at the entrance of the environment and other agents move to get into formation. 
All the techniques above do not incorporate our game rules such as, boundary condition, dynamically changing length of the chain, chasing and surrounding a competitive prey. We develop Catch-Catch and Chain Catch game as multi-agent system where the world of game is
set as a rectangular grid model and each agent is modeled as an autonomous agent. We use cost model to drive strategies that have implicit formation tactics, game rules such as boundary cross, and collision avoidance embedded into them. Using this solution, we have developed four strategies for Escapees and two techniques for chain formation that is presented in section 4. We also performed a number of empirical experiments to analyze and compare performance of the strategies. We used production quality robots to implement the game play in physical environment and showing viability of our solution.

\section{Agent Strategies}
We use a cost model to develop strategies for each of Escapee, Catcher and Chain. The cost functions are a means to estimate the effectiveness of a move or decision taken by the agent. Lesser the cost of a cell, better it is for the agent to move into it.
\subsection{Catchers strategy}

The game starts with one agent playing as Catcher trying to chase and Catch one of the Escapees. Let the coordinates of Catcher ($C_x$, $C_y$) and coordinates of Escapees be $E$=$\left\{(x_1,y_1),(x_2,y_2),………..(x_{(n-1)},y_{(n-1}))\right\}$ where n is equal to the number of agents in the game. 
Then let the function of \textit{Cost(x,y)} (where, \textit{(x,y)} are coordinates of cell) for Catcher strategy be
\begin{equation}
 Cost_{nd}(x,y) = \sqrt{(x-E_x)^2 + (y-E_y)^2}
\end{equation}
Here $E_x$ and $E_y$ are the coordinates of the Escapee that has minimum distance $D_c$ from the Catcher.
\begin{equation}
 E_{D_c} = \arg\min_{i}(Distance(Catcher\;c, Escapee\;e_i))
\end{equation}
 At each cycle, Catcher computes Euclidean distance from all the escapees, and the minimum distance among them takes the Catcher towards nearest Escapee to catch the Escapee.
\subsection{Escapees strategy}
Main criteria for Escapees$'$ strategy is the distance to the Catcher or chain that is chasing them. Distance to Catcher can be determined based on the coordinates of cell where Catcher resides but however to determine effective distance to chain a \textit{Representation Point} of chain needs to be decided. \textit{Representation Point} is chosen based on the member of chain that is nearest from the Escapee. Let minimum distance $D_{ch}$ from the Escapee be-
\begin{equation}
D_{ch}= \min_{i}(Distance(Escapee\;E,Chain\;Members\;Ch_i  ))       
\end{equation}
 Note that the \textit{Representation Point} varies for each Escapee and it also changes at each cycle of the game, as relative positions of the agents change at each cycle.
\subsubsection{Maximize Distance or Naive Strategy}

With all the notations above keeping intact, consider a function.
\begin{equation}
 Cost_{md}(x,y) = MAXCONSTANT - Catcher\;Distance
\end{equation}
\begin{equation}
 Catcher\;Distance = \sqrt{(x-C_x)^2 + (y-C_y)^2}
\end{equation}
With \textit{(x,y)} as coordinates of the cell and $(C_x , C_y)$ as the coordinates of Catcher or \textit{Representation Point} of chain. \textit{Catcher Distance} (CD) is Euclidean distance from cell to the chain and Maxconstant is set to a value greater than maximum possible distance between two cells in game field. 

The cost is maximum at the position of the cell where
Catcher or the chain$'$s \textit{Representation Point} itself is located. And as we go far from the Catcher/chain the \textit{Catcher Distance} increases, thus decreasing the Cost. Since aim of an agent is to move to the cell with minimum cost, this cost function moves it away from Catcher/chain and to evade the chain. 
\subsubsection{K circle Strategy}

We extend our strategy for the Escapees to multiple criteria. It not only depends upon the \textit{Catcher
Distance} (CD) but also distance to fellow Escapees. Consider a cost function,
\begin{equation}
 Cost_{k}(x,y) = |K - Catcher\;Distance|
\end{equation}
Where \textit{K} is the safe distance we want all Escapees to maintain from Catcher or Chain$'$s \textit{Representation Point}. This cost function becomes zero (minimum) only at the cell points where \textit{Catcher Distance} is equal to \textit{K}. The idea is to provide multiple Escapees as  catchable preys to Catcher/chain for it to decide which Escapee to pursue. Now, from the Escapees point of view, we further want to complicate things for the chain by limiting chain's movable area and surround it completely with Escapees. Also, as the distance between Escapees decreases, probability of their collision also increases, hence we need to introduce additional cost due to neighbor Escapees that are substantially close to it.

Consider a new cost function

\begin{equation}
Cost_{kc}(x, y) = 
\begin{cases}
    |K - CD| + NSD - NND,& \textbf{if} \\\mathbf{(NND < NSD)}\\
    |K - CD|, & \textbf{otherwise}
\end{cases}
\end{equation}
 \textit{Nearest Neighbor Distance (NND)} is the distance from cell to the closest fellow Escapee. 
\begin{equation}
\begin{split}
NND = \min_{i\neq{j}}(Distance(Cell,Escapee\;e_i))
 \end{split}
\end{equation}
And \textit{Neighbour Safe Distance (NSD)} is the minimum safe distance an Escapee should maintain from other Escapees and j is the Escapee whose move is being calculated. Using this cost function Escapee tends to maintain safe distance from fellow Escapees while achieving a spread among themselves and maintaining \textit{K} distance from the Chain, leading to a circle like formation with a radius equal to \textit{K}. The value of K is set to enable formation of freely moving \textit{K} circles in the region.
Value of \textit{Neighbour Safe Distance (NSD)} is decided based upon number of agents and their diameter. Considering the diameter constant, if number of agents are less, greater the value of \textit{Neighbour Safe Distance}, better and circular is the spread around the Catcher, whereas if the number of agents are significantly high, then a small value of safe distance helps form a better formation and avoid collision.    

\subsubsection{K circle Strategy with Rotation}

We enhance K circle strategy by making the Escapees rotate around the Catcher or \textit{Representation Point} of chain when it reaches the K circle. Consider cost function when such condition is met:
\begin{equation}
\begin{split}
 Cost_{kr}(x,y) = Distance\;to\;Rotation\;Point, \\
\mathbf{ if (K - K_2) \leq Catcher\;Distance < (K + K_2)}
 \end{split}
\end{equation}
The cost function remain same as in equation 7 in all other cases.
\begin{equation}
 Distance\;to\;Rotation\;Point = \sqrt{(x-R_x)^2 + (y-R_y)^2}
\end{equation}
Where \textit{(x,y)} are coordinates of the cell and ($R_x$,$R_y$) are the coordinates of the rotation point such that,
\begin{equation}
\begin{gathered}
R_x = C_x + K * \cos(\theta + d\theta),\\
R_y = C_y + K * \sin(\theta + d\theta)
\end{gathered}
\end{equation}
Where ($C_x$,$C_y$) are coordinates of Chain's \textit{Representation Point}/Catcher as center and \textit{K} is the radius of the K-circle. $\theta$ is the angle the Escapee under consideration makes with respect to the Catcher/chain at that instant in the game field. $d\theta$ is the angle with which we aim to rotate the Escapee. \textit{$K_2$} is a parameter used to define the range of K radius. It is kept as unit cell in our simulations.

\begin{figure}[!tbp]
  \centering
  \begin{minipage}[b]{0.2\textwidth}
    \includegraphics[width=1.5in, height=1.7in]{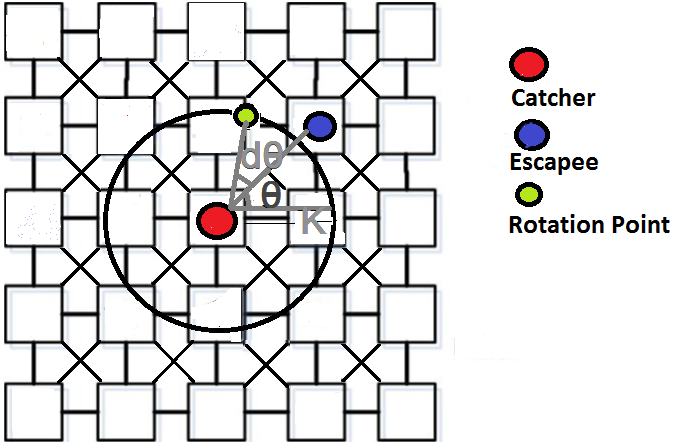}
    \caption{Depiction of Rotation point in K circle strategy with rotation.}
    \label{img:4.2}
  \end{minipage}
  \hfill
  \begin{minipage}[b]{0.2\textwidth}
    \includegraphics[width=1.5in, height=1.7in]{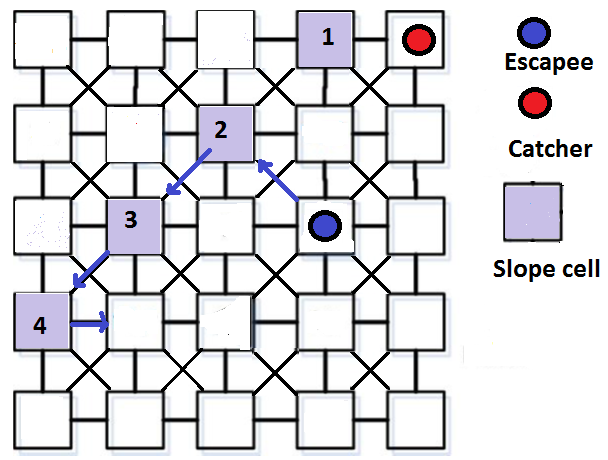}
    \caption{Example of Motion of Escapee on the slope with a slope at North West corner.}
    \label{img:4.3}
  \end{minipage}
\end{figure}

In the equation 9, cost function is equal to Euclidean distance of a cell from a \textit{Rotation Point} when the Escapee is lying on the K circle or within range of \textit{K}( \textit{K} $\pm$ \textit{$K_2$}). As the distance of a cell decreases from given \textit{Rotation Point}, cost also decreases and the function becomes zero at the given point. Since an
agent$’$s aim is to stay in a cell with minimum cost, this function leads the Escapee to move towards a cell that is closest to such a point on the K circle (with Catcher as center) which makes $d\theta$ angle with current position of the Escapee (see Figure \ref{img:4.2}). Ultimately the strategy makes all the escapee spread among themselves forming a circle with radius equal to K and also rotate around it.

% % % % % % % % 
\subsubsection{Sliding Slope strategy}
This strategy is another extension to the K circle strategy with cost function in equation 7. Note that, Escapees are most prone to get captured at the corners of the game grid, because two sides of the field are bounded and therefore, limiting agent$'$ moves. To avoid a trap near the corner we introduce a virtual slope at four corners of the arena through which an Escapee can slide. Slope is an imaginary diagonal edge near the corners of game arena, whose path an Escapee can take to escape from chain.

In this strategy Escapee moves with cost function given in equation 7, except when it strikes
one of the grid cells part of a sliding slope. In that case its Legal Moves are restricted to only the cells that are along the slope it is on. Once it lies on the slope, it chooses to go in direction of the slope end that is farther from the chain (see Figure \ref{img:4.3}).
\begin{table}
\begin{tabular}{ |p{2cm}|p{6cm}| }
\hline
\textbf{Strategy name} & \textbf{ Cost function Description}  \\
\hline
Maximize distance (Naive) &  Escapees maximize their distance from Catcher and gather near boundaries of arena \\
\hline
K circle & Escapees surround Catcher in a circle with radius equal to ``K'' maintaining \textit{NSD} from nearest Escapee \\
\hline
K circle with rotation & Escapees surround Catcher with in a circle and rotate around it \\
\hline
Sliding slope & Escapees form a K circle and slide towards the end of the sliding slope that is farther from Catcher or chain \\

\hline
\end{tabular}
\caption{ Summarizing strategies for Escapees.}
\label{table:4.1}
\end{table}
Table \ref{table:4.1} summarizes all strategies discussed in section above. It describes cost function's equations in words and briefly explains their effect on formation of Escapees.
\subsection{Chain's Strategy}
Member agents of the chain have dual objectives- (i) Catch an Escapee (ii) maintain chain formation. We have designed following strategies for chain members keeping the two objectives under consideration.
\subsubsection{Tagging Method}
\textbf{}\\
\textbf{\textit{ Case 1: Catch by ends of chain}}\\
Consider three terms for this strategy-(i)\textit{Leader of chain} (Chain member that chases the Escapee), (ii) \textit{leader agent} (any agent of the game, which is being followed by a chain member),(iii) \textit{Tagged member}, Chain member which follows or moves in sync to its \textit{leader agent} . We first find the \textit{Leader of chain} by finding the corner member that is closest to the nearest Escapee in the game field. Consider a game of \textit{n} agents with \textit{m} agents member of chain with coordinates Ch=$\left\{(a_1,b_1),
(a_2,b_2),………...(a_m,b_m)\right\}$ and Escapees  E=$\left\{(x_1,y_1),(x_2,y_2),
………..(x_{(n-m)},y_{(n-m)})\right\}$. 

\begin{figure}[!tbp]
  \centering
  \begin{minipage}[b]{0.2\textwidth}
    \includegraphics[width=1.5in, height=1in]{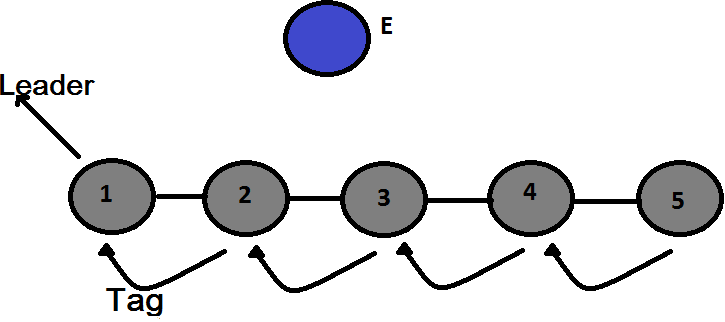}
    \caption{An example of Chain Tag}
    \label{img:5.1}
    \end{minipage}
  \hfill
  \begin{minipage}[b]{0.2\textwidth}
    \includegraphics[width=1.5in, height=1in]{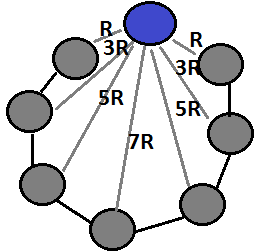}
    \caption{Chain concave formation around Escapee with Variance function.}
    \label{img:5.3}
  \end{minipage}
\end{figure}
Once the \textit{Leader of chain} is assigned; it also becomes the \textit{leader agent} for the member next to it. This member next to \textit{Leader of chain} gets tagged to its \textit{leader agent} such that the tagged member always moves to the cells adjacent to the \textit{leader agent}. This tagged member then becomes \textit{leader agent} for the agent next to it in the chain. Such a tag propagates until the other end of the chain. Consider Figure \ref{img:5.1} to understand the tag.\\ 

Now, consider the cost function of this strategy.
\begin{equation}
Cost_{tm}(x,y) = |R_{safe} -D_l|
\end{equation}	
And,
\begin{equation}
D_l = \sqrt{((x-l_x )^2+(y-l_y)^2 )}
\end{equation}
Where \textit{(x,y)} are the coordinates of the cell. And for the Cost(x,y) for the \textit{Leader of chain}, $ R_{safe}$  is the safe distance from the \textit{Leader of the chain} to the Escapee it is chasing (it is kept equal to diameter of the agent in our simulations) such that it just touches the Escapee. And $D_l$  is the Euclidean distance of a cell (x,y) to the \textit{leader agent} $(l_x,l_y)$ which in this case is the Nearest Escapee $(E_x, E_y)$, the Leader is chasing.

Now, for the cost(x,y) for the other members of the chain $R_{safe}$ is the safe distance between neighbor members of the chain. And $D_l$ is the Euclidean distance of a cell (x,y) to the \textit{leader agent} $(l_x,l_y)$ of the chain member to which it is tagged. For a chain member $Ch_i$ ($a_i,b_i$) its \textit{leader agent} $(l_x,l_y)$ is, 

\begin{equation}
\begin{split}
(l_x,l_y)=
\begin{cases}
(a_{(i-1)}, b_{(i-1)}), &\textbf{if} \\ \mathbf{Leader\;of\;the\;chain\;is\;the\;last}\\\mathbf{member\;of\;the\;Chain\;(a_1,b_1)} \\
(a_{(i+1)}, b_{(i+1)}),   &\textbf{if}\\ \mathbf{Leader\;of\;the\;chain\;is\;the\;last}\\\mathbf{member\;of\;the\;chain\;(a_m, b_m)}\\
(E_x,E_y), &\textbf{if} \\ \mathbf{Ch_i\;is\;Leader\;of\;chain}\\
\end{cases}
\end{split}
\end{equation}
 The cost function equation shows that this function tries to maintain a chain member$'$s distance to $R_{safe}$ from its leader as it attains its global minimum (equal to zero) at $D_l$ equal to $R_{safe}$. $D_l$ for the Leader of the chain is the distance from nearest Escapee it is chasing. This condit
 ion becomes similar to Catcher$’$s strategy discussed earlier where the \textit{Leader of chain} tries to minimize its distance to Escapee. To keep the chain in sync with motion of the Leader, other members try to maintain distance $R_{safe}$ from their leader agent to which they are tagged in direction of the Leader of the chain (see Figure \ref{img:5.1}). \\
 \\
 \textbf{\textit{Case 2: Catch by any chain member}}\\
In this case also functionality of Tagging method remains same except that there are more choices for the Leader. Leader is assigned to the chain member that is closest to the nearest Escapee in the game field.
 Here all the members left to the Leader get tagged to the members right to them up till the Leader and all the members right to the Leader get tagged to members on their left in direction of the Leader. 

Once the agents are tagged, the cost function in equation 12 is applied to each of chain member and they move to the cell with minimum cost. Therefore, we get the Leader moving in direction to the nearest Escapee and other members in direction to their leader members with safe distance $R_{safe}$. This way we are able to get a synchronized motion of chain in direction to its target Escapee.
\subsubsection{Variance Method}
In this method, we use the concept of Variance function that decides the distance a chain member should maintain from the Escapee.\\
\\
\textbf{\textit{Case 1: Catch by ends of chain}}\\
Here, Leader of the chain is chosen based on its distance from the nearest Escapee. Since in case 1 only corner members can catch an Escapee, we find the corner member that has minimum distance from the Escapee it has to chase. Escapee to be chased at each cycle of the game is chosen to be one with minimum distance from the Leader of the chain compared to other Escapees.

 In our variance function, we try to surround an escapee by forming a loop like structure around it as shown in Figure \ref{img:5.3}. We set variance distance of two ends of the chain to $R_{safe}$ as the corners need to catch the Escapee. Then increment variance value from corner to the middle member. If chain has \textit{n} (1 to n) members then variance value of $i^{th}$ member will be
\begin{equation}
\begin{split}
Variance[i]=
\begin{cases}
variance[i-1]+R, &\textbf{if}\\ \mathbf{i\in{(1,m/2]}}\\
Variance[i-1]-R,  &\textbf{if}\\ \mathbf{i \in{(m/2,m)}}\\
R_{safe},                &\textbf{if}\\ \mathbf{i\in{{1,m}}}\\
\end{cases}
\end{split}
\end{equation}
Where R is the radius of the agents.  \\
Consider the following cost function,
\begin{equation}
Cost_{ve}(x,y) = |R_e -D_e|
\end{equation}
Where $R_e$ is the variance distance from Escapee,($R_e$=Variance[id]) And $D_e$ is the distance of cell \textit{(x,y)} from the Nearest Escapee $(E_x, E_y)$. As we can see in equation, this cost function [16] assures a Chain member to maintain a distance $R_e$ from the Escapee, the chain is chasing. The cost function attains its minimum at $D_e$ equal to $R_e$, where $R_e$ is given by the variance function defined above. However this function does not assure maintenance of chain formation. Therefore, the cost function is modified to the following,
\begin{equation}
\begin {split}
Cost_{vh}(x,y)=
\begin{cases}
|R_e-D_e |  &\textbf{if}\\ \mathbf{(r_1< D_c <r_2 )}\\
|R_e-D_e| + |R_c-D_c | \\&\textbf{otherwise}  
\end{cases}
\end{split}
\end{equation}
Where $R_c$ is the safe distance between two neighbor members of the Chain; it is kept as average of $r_1$ and $r_2$ discussed in section 2. And $D_c$ is the Euclidean distance of cell \textit{(x,y)} from the neighbor Chain member. If a chain member does not lie within range of $r_1$ and $r_2$ from its neighbor, it is provided with additional cost $|R_c-D_c |$ of staying outside the range. Hence the two functions combined try to move the chain in such a way that it surrounds Escapees with given Variance and maintain safe distance among each other.\\
\\
\textbf{\textit{Case2: Catch by any chain member}}\\
Here, Leader is that has minimum distance from the closest Escapee. \\
Since in this case a Leader can be a member in between the chain as well, it has to have smaller value of Variance distance from the Escapee unlike the variance in case-1. If index of the Leader in the chain is LeadIndex and number of agents in the chain are \textit{m} then  
 \begin{equation}
 \begin{split}
 Variance[i]=
 \begin{cases}
 R_{safe} &\textbf{if}\\ \mathbf{i = LeadIndex}\\
 Variance[LeadIndex] + \\|LeadIndex - i|*R &\textbf{otherwise}
 \end{cases}
 \end{split}
 \end{equation}
 We organize the chain members such that the Leader is closest to Escapee and others are farther forming concave structure in direction opposite to the Escapee. Same cost that in equation 17 can be used here as well. With Dc as distance from the neighbor in direction of the Leader. \\
 Table \ref{table:5.1} summarizes all strategies discussed for chain in this section. \\

\begin{table}
\begin{tabular}{ |p{1.2cm}|p{1.7cm}|p{5cm}|  }
\hline
\textbf{Strategy name} & \textbf{Leader} &  \textbf{Description}  \\
\hline
Tagging method (Case-1) & Nearest corner of chain from Escapee & Leader moves towards nearest Escapee. Other members tag themselves to their neighbour in direction of Leader and move to the cell closest to it\\
\hline
 Variance method (case-1) & Nearest corner of chain from Escapee & All members try to attain different \textit{variance} distances from Escapee to surround it while maintaining safe distance \textit{$R_c$} from their neighbours  \\
\hline
Tagging method (case-2) & Nearest chain member from Escapee & Leader moves towards nearest Escapee. Other members to its left and right tag themselves to their neighbour in direction of Leader \\
\hline
Variance method (case-2)   & Nearest chain member from Escapee &  All members try to attain different \textit{variance} (as defined for case-2)  distance from Escapee while maintaining \textit{$R_c$} from neighbours  \\

\hline
\end{tabular}
\caption{ Summarizing strategies for chain}
\label{table:5.1}
\end{table}

We handle boundary condition by adding cost value of infinity to the cells on the arena that are at the boundary edges of the game field and where another agent is already residing. Once such cells are filtered out only then the cost functions are applied to calculate the cost of other legal cells. This ensures that agents do not cross the boundary walls and still adjust their formation and movement accordingly.
% end the environment with {table*}, NOTE not {table}!

\section{Robot Simulation}

 We use production quality Robotic platforms Fire Bird- V ATMEGA2560 for this purpose that are designed and built by Nex robotics \cite{10}. The communication is done through Xbee API module. Our robotic setup does not have localization mechanism therefore, we implement virtual localization through communication. These robotic agents are similar in terms of size, speed (same and constant) and behavior. To establish communication among the agents through a coordinator we follow a message protocol. The coordinator sends an initial information message packet to all the agents that contains unique ID to each robot, their initial \textit{x} and \textit{y} coordinates and \textit{CatchMode}. Users have to place the robots onto the specified starting location to begin the game with. Once the game begins, the robots computes the best move possible depending upon information it has about other agents using the Strategy Engine module. These moves are any one of Legal Moves discussed earlier. The Move is then processed by the Motor Control module, which converts the motion command into corresponding control signal that is fed into Motor driver through GPIOs. We also use a control algorithm to maintain and balance our robots as autonomous self-balancing four wheel robot based on the PID controller \cite{55}.
We have six robots; we implemented each of Catcher, Escapee and Chains algorithms on these robots in different scenarios and examined the performance.

\begin{figure*}[htb]
 \includegraphics[width=6.5in, height=2.9in]{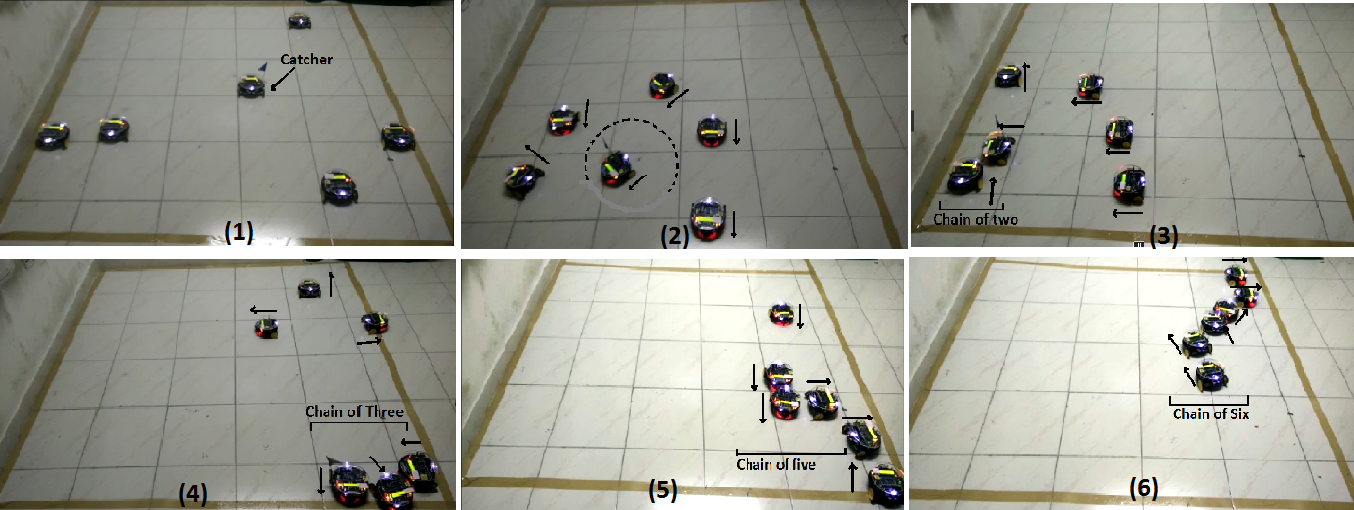}

 \caption{Robot simulation of Chain game with chain$’$s Tagging method and Escapees moving with K circle strategy. (1) Initial condition. (2) K circle around Catcher. (3) First Catch (4) Chain of three (5) Chain of five. (6) Chain completion}
 \label{img:5.8}
\end{figure*}
Consider an example in Figure \ref{img:5.8} Here we take game field grid of size 220x220 cm where our robots$’$ diameter is 16 cm. There are six robots playing the game with middle player with flag playing as Catcher (part (1) of figure). Black arrows in the snapshots depict the direction of motion of robot.\\
Here we apply cost function discussed in equation 7 for the Escapees. Therefore, they attempt to form a circle-like formation around the Catcher as seen in part (2) of the figure. We observed that more the number of players, fuller is the formation of circle. As we can see in part(3), after the first Catch,the number of Escapees reduce down to four and when the chain resides near the corner of the game field, Escapees form a K-arch like formation around the chain. When two players satisfy the condition of Catch discussed earlier, we consider them to form a chain. However a physical touch between robots is not encouraged in our implementation, as it leads to collision among them and that disturbs their direction of motion in a continuous game. Hence when two are in vicinity of each other as shown in part (3), we consider it to be a Catch. The chain expands every time a Catch occurs (Figure \ref{img:5.8}, part (3,4,5,6)). Here chain moves with Tagging approach. We observe synchronized move of robots in the chain in each round of Catch. Consider figure \ref{img:5.8} part (6)) where directions of each robot shows how the last Escapee reaches and oscillates near North East corner of the grid while escaping from chain, Leader of chain chases it to the corner and other chain members align themselves to the Leader meeting boundary condition as well as chain constraints forming a chain of six. Live videos of some of the game play experiments can be found in reference section \cite{12}.	\\
\section{Experimental Results}

While results of robot implementation are promising, we further analyze the game with increasing number of agents and varying different parameters with the help of a simulator. We have built a front-end simulator to design and experiment with various Chain Catch game strategies. The simulation in simulator takes place as a series of Chain-Catch rounds and cycles. Input is taken from the user for number of agents, initial \textit{x}, \textit{y} coordinates and \textit{CatchMode} of each agent. Once the role of each agent is decided, game starts and at each cycle the corresponding move function of each agent is called and thus updating its position on the game field. Each and every change in
the environment is noted by the coordinator at each cycle and takes appropriate actions based on the game rules embedded into it such as detecting Catch, boundary cross etc. The simulation view panel also notes these changes and renders the field view periodically. 
\begin{figure}[htb]
 \includegraphics[width=3.5in, height=3in]{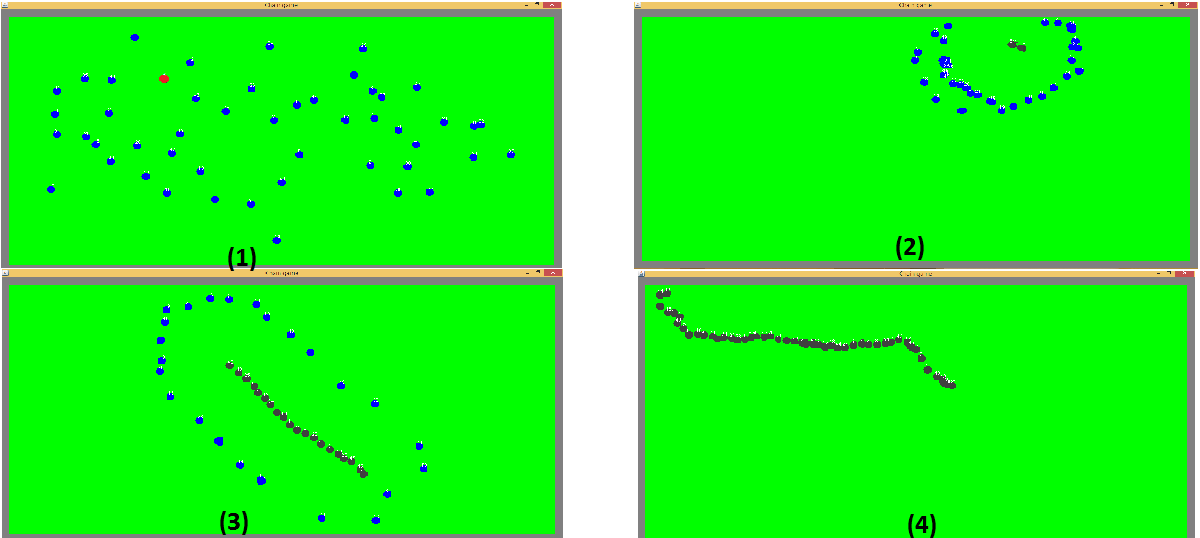}
 \centering
 \caption{ Simulation of chain using Tag approach and Escapees moving with strategy of K circle. (1) starting position, (2) Chain of two (3) elliptical like formation of escapees around chain (4) synchronized chain completion }
 \label{img:5.4}
\end{figure}
 Figure \ref{img:5.4} is an implementation of chain$’$s strategy using cost function shown in equation 12. It demonstrates a simulation of Chain game with 50 agents, where the chain moves with Tagging method whereas the Escapees move with K circle strategy. In these simulation we keep the value to K to be one fourth of arena$'$s vertical length. And value of \textit{Neighbour Safe Distance (NSD)}, is varied from two times Agent Diameter to four times Agent Diameter. For the chain$'$ movement, value of $R_{safe}$ is kept as 3*radius of the agent in our simulations. And the Game Over condition is kept as 3000 steps (an agent takes 300 steps to walk through periphery of arena). 
\begin{figure}[htb]
 \includegraphics[width=3.5in, height=3in]{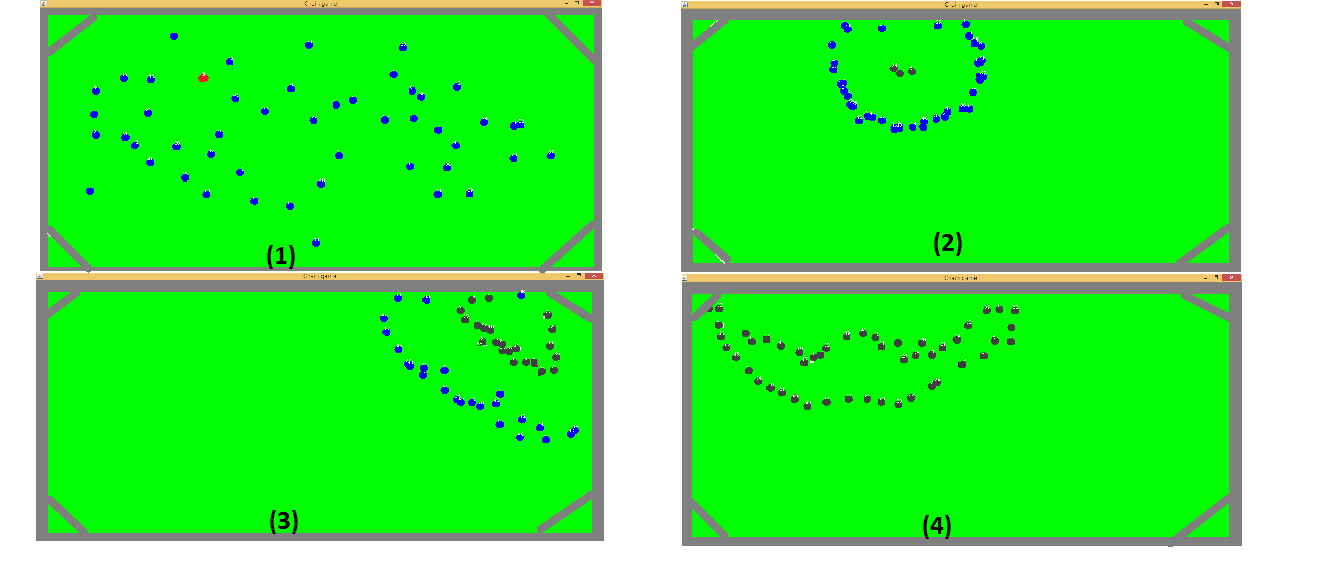}
 \centering
 \caption{ Simulation of chain using Variance approach and Escapees moving with strategy of Sliding Slope (1) Game start, (2) K circle (3) Chain surrounding Escapee (4) Complete chain formation}
 \label{img:3.4}
\end{figure}
Figure \ref{img:3.4} is an implementation of cost function shown in equation 17. Figure \ref{img:3.4} is an example of the same, where chain members move with variance explained in equation 15 and are able to surround an Escapee near corner (Figure \ref{img:3.4} Part (3)) and Escapees move with the strategy of Sliding Slope.

\begin{figure}[htb]
 \includegraphics[width=3.5in, height=2.3in]{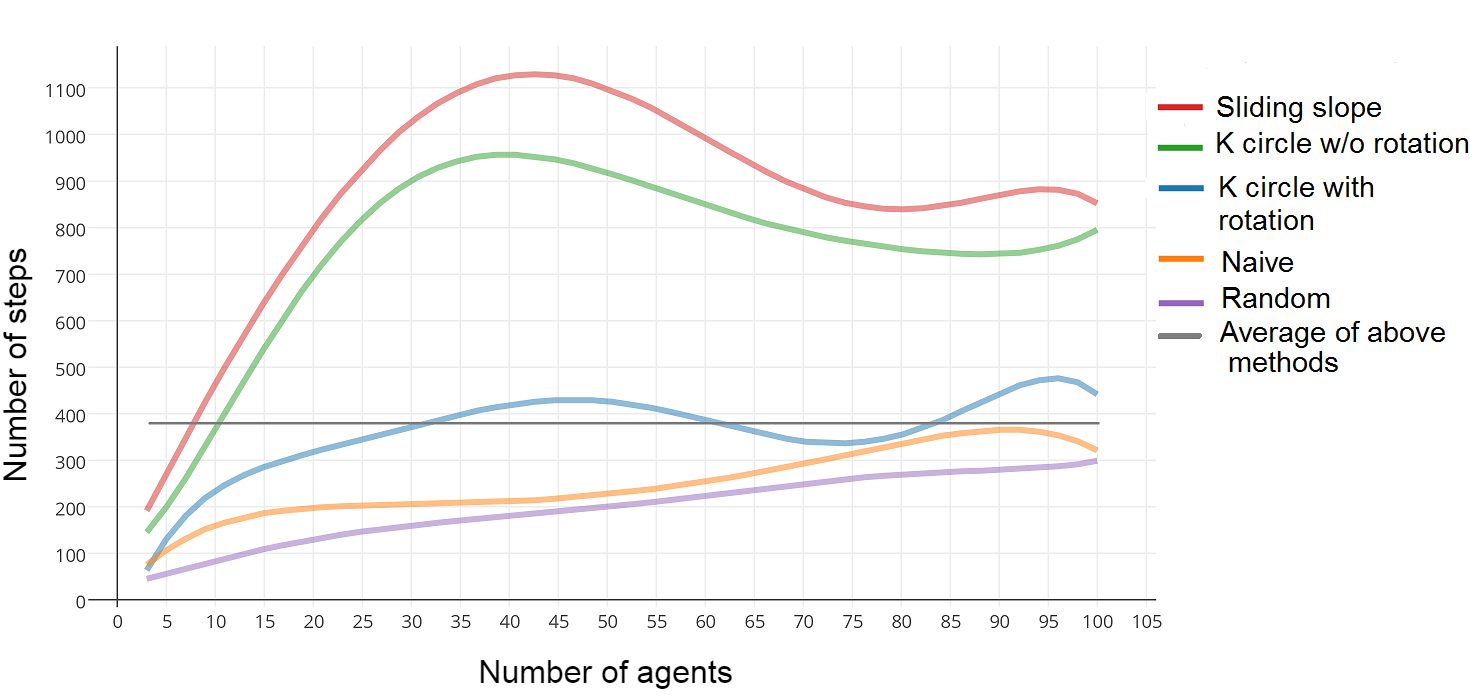}
 \centering
 \caption{Escapees' strategies performance against chain Tag method.}
 \label{img:6.3}
\end{figure}

We have implemented a set of strategies for each type of agents- Escapee, Catcher and Chain in the
game that have been discussed in previous sections. We also implemented random movement of Catcher, Escapee and Chain to show how other strategies work better than cases where an agent chooses to move randomly. However we need well defined parameters to understand each strategy`s performance and to compare them. We use the parameter of Total time ($T_c$) to study the performance, which is the time taken till the last Escapee gets caught by the chain. It also denotes
number of steps taken by the chain for complete chain formation or before Game Over. From chain$’$s perspective, chain strategy that takes minimum $T_c$ to finish the game is considered to have best performance and from Escapee$'$s perspective Escapees$’$ strategy that takes longest time $T_c$ to finish the game is considered as best strategy. We performed over 100 such empirical experiments, with varying number of agents from 3 to 100, and different starting locations of agents in each case. Figure \ref{img:6.3} plots $T_c$ values for those 100 experiments. On X axis number of
agents participating in game are plotted whereas Y axis denotes number of steps taken to finish the the game ($T_c$) corresponding to \textit{n}(number of agents playing in a game).  Plot plots $T_c$ values for each of Escapee strategy - Naive (orange), K circle (green), K circle with rotation (blue), Sliding slope (red) and random movement (purple) when chain plays with Tagging method. 
\begin{figure}[htb]
 \includegraphics[width=3.5in, height=2.3in]{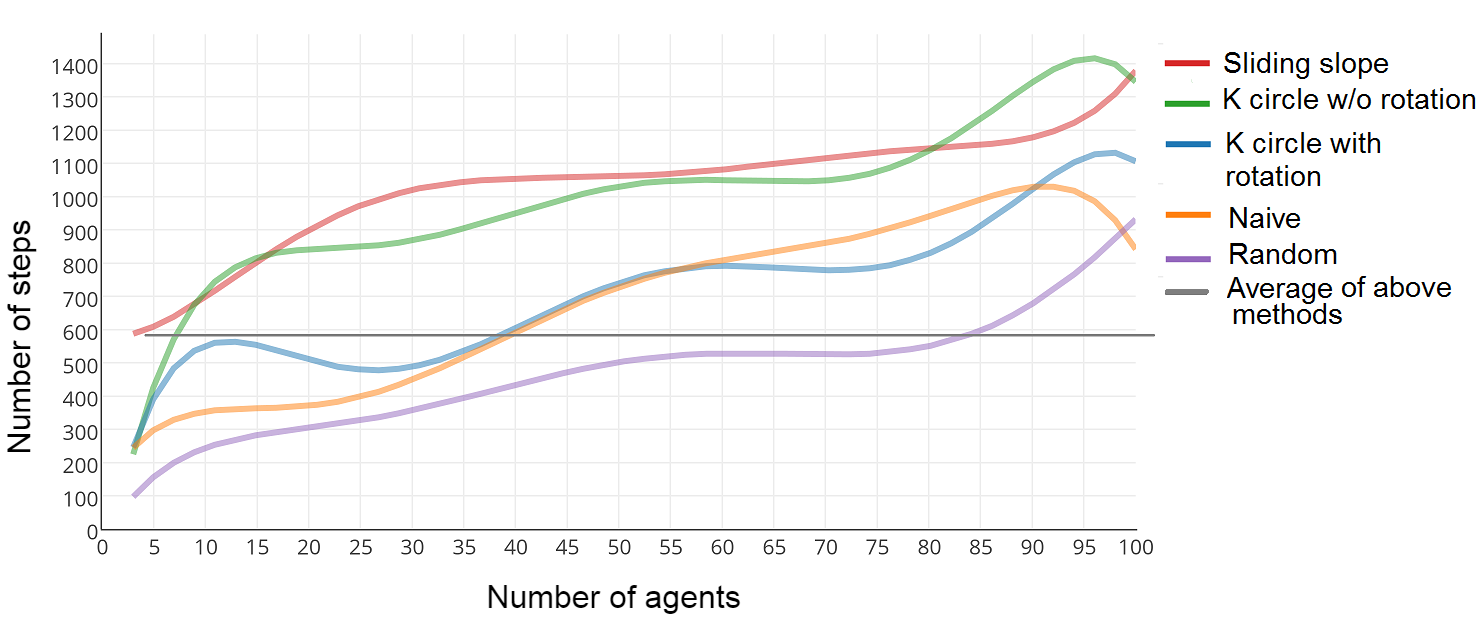}
 \centering
 \caption{Escapees' strategies performance against chain Variance method.}
 \label{img:6.4}
\end{figure}
Sliding slope (red plot) strategy has the best performance as it takes highest number of steps to finish the game in almost all experiments. Whereas Maximize distance and random movement strategy have relatively poor performance. The gray line shows an average of total number of steps ($T_c$) in all 100 experiments when chain uses Tag method. Figure \ref{img:6.4} shows performance of each of Escapee strategies when implemented against chain$’$s Variance method. Here too overall comparison leads to same order of performance of the Escape strategies as in case of Tag method discussed before. 
 \begin{table}[htb]
 \includegraphics[width=3.5in, height=2in]{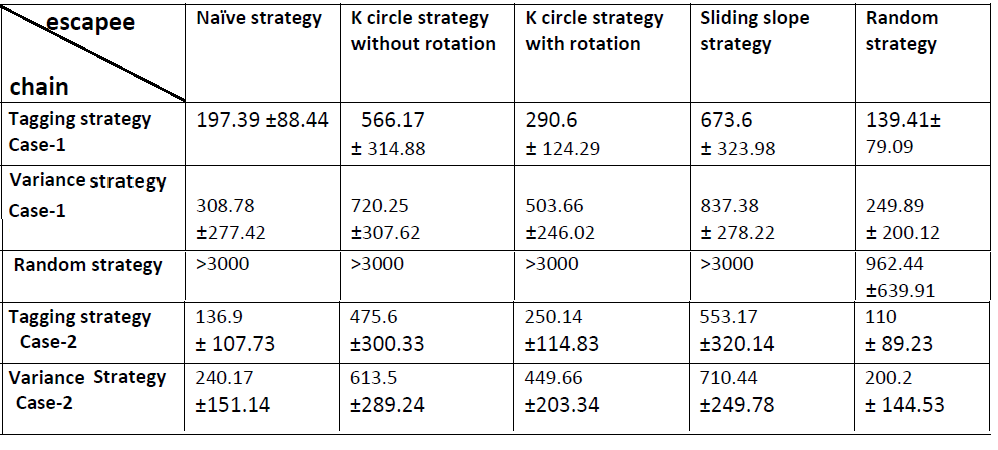}
 \centering
 \caption{Average number and standard deviation of steps ($T_c$)}
 \label{Table:6.4}
\end{table}
We have discussed in strategies section that there are two cases of Chain Catch possible. Case 1, where only corner members of the chain participate in Catch. Case 2, where every member of chain is allowed to Catch an Escapee. Chain$’$s performance differs in each case. Our each experiment had 25 runs for each combination of Escapee and chain$’$s strategy, leading to overall 2500 runs. Transition table shown in Table \ref{Table:6.4} gives overall analysis of game results in terms of average number of steps taken for the complete chain formation. Cells in the table also gives standard deviation for each case. Left to right columns in the table represent Escapees strategies and top to bottom rows represent strategies for the chain. The data $>$3000 implies that the strategy is insignificant; as it takes more than maximum possible steps (3000) to finish the game.

Our game simulations, robot implementation and experiment results suggest that both of our chain strategies are able to achieve successful chain formation. Since we had only six robots - we showed how our strategies work and provide the similar behavior to what the strategy expects with robots. We implemented Escapee$’$s random movement and Naive approach as a benchmark to compare newly introduced strategies. All the comparison plots of $T_c$ and transition table suggests that each version of our K circle strategy performs better than the benchmark. The order of performance is\\
\textit{Random Movement $<$ Maximize distance (Naive) $<$ K circle with rotation $<$ K circle $<$ Sliding Slope}\\
 And Chain$'$s movement methods work in following order of performance-\\
\textit{Random movement$<$ Variance method (case-1)$<$Variance method (Case-2)$<$ Tagging method (Case-1)$<$ Tagging method (Case-2)}\\
From Escapee$'$s perspective, its objective is to delay a Catch as long as possible and delay overall chain completion. Performance of our Escapee strategies against a random moving Catcher (third row of Table \ref{Table:6.4}) proves them meeting this objective. Escapee$’$s random movement is last in order of preference because here they move without an incentive of evading themselves from a Catch. Maximize distance or naive approach is a strategy inspired from Korf$’$s method \cite{34}, moves Escapees with clear incentive to move farthest from Catcher, therefore, it makes Escapees perform better than randomly moving Escapees. But it does not include any criteria of cooperation from other fellow Escapees. Therefore, it leads many Escapees to gather at one place (usually near corners) and hence enabling chain to Catch one Escapee after another within very few steps. This problem is addressed in cost function defined for K circle, K circle with rotation and Sliding slope methods. Here Escapees spread among themselves while maintaining safe distance from Catcher and hence their performances are better than Maximize distance approach. However, when Escapees try to rotate around Catcher by moving in direction of \textit{Rotation Point}, they sometimes end up moving towards walls. This causes more early Catches in game compared to original K circle strategy. This is, why K circle strategy performs better than one with rotation. Introduction of virtual slope further decreases probability of an Escapee getting stuck near the corner and hence gives best results among all as seen in transition table. Escapees$’$ performance also depends upon strategy chosen by the opponent ie. chain as seen in
transition table.

Chain can move as per Tagging approach discussed earlier. This approach causes a
synchronized movement of all chain members. As here only one member of chain acts as Leader and
moves in direction of Escapee, rest simply align to it. Therefore, it performs the best as suggested in Table 1. Variance method makes all the chain members move in direction of Escapee. Its game simulation seems more like human game where, in a long chain, different members of chain try to move into different directions and hence cause strain in the chain formation. And since we reject a Catch occurred during chain break condition, overall time increases to get complete chain formation. \textit{Chain$’$s random movement strategy works only when Escapees also move randomly}. Hence having overall worst performance. In case-2 (Catch by any chain member) of chain movement, chain attains an additional functionality of being able to Catch by members in between the chain. Numbers shown in transition table suggest that this functionality definitely improves performance of chain in each possible case. That is, average number of steps to finish the game ($T_c$) in case-2 (Catch by any chain member) is less compared to case-1 (Catch by ends of chain) with both chain strategies against all five Escapee strategies.  Though an individual agent$’$s performance in Chain-Catch game depends upon many factors like starting positions of each agent, relative distance from initial Catcher, number of agents and opponent$’$s strategy but the overall analysis suggests that choice of strategy affects overall performance most significantly driven by their comparative order given above.

\section{Conclusions}
We built a Multi-robot system where robotic agents are capable to play Catch-Catch and Chain Catch. We implemented the system both
as simulation framework and in physical environment with real robots. Movement of robots in a Catch-Catch game or in formation of a chain requires coordination amongst multiple robots, which makes our framework useful as a ``search and rescue'' robot system. An example of Chain Catch is where, to trap a terrorist the robots might have to form a chain and move in a coordinated fashion or even surround it with a circular formation as done by our chain and Escapees.

Our results show that Sliding slope strategy is the best strategy for Escapees whereas Tagging method is the best method for chain$’$s movement in Chain Catch. As a part of future work, the simulator can be made more generic in form of a simulator package
library that can be used by any MAS researcher to develop his own pursuit domain strategies and use it as test-bed to evaluate them. And for robot implementation an external camera can be added to the setup to achieve localization in robots by detecting other agents, walls and obstacles by themselves.

%Acknowledgements are optional

%\balancecolumns
%
% For AAMAS-2016, as references are unlimited but appendices must fit within
% 8 pages, the References section must come after the appendices (if any)

\bibliographystyle{abbrv}
\bibliography{sigproc}  % sigproc.bib is the name of the Bibliography in this case
% ACM needs 'a single self-contained file'!
%\balancecolumns % GM June 2007
% That's all folks!
\end{document}